\documentclass[%
 reprint,
superscriptaddress,
 amsmath,amssymb,
 prl,
]{revtex4-2}

\usepackage{graphicx}
\usepackage{xcolor}
\usepackage{physics}
\usepackage{amsfonts}
\usepackage{amsmath}
\usepackage{booktabs}
\definecolor{red}{rgb}{1,0,0}

\usepackage{dcolumn}
\usepackage{bm}
\usepackage{hyperref}
\hypersetup{
colorlinks=true,
linkcolor=black,
filecolor=black,  
citecolor=black,
urlcolor=black,
}

\usepackage{pdfpages} 
\usepackage{pgffor} 

\makeatletter
\AtBeginDocument{\let\LS@rot\@undefined}
\makeatother

\def\supplementfilename{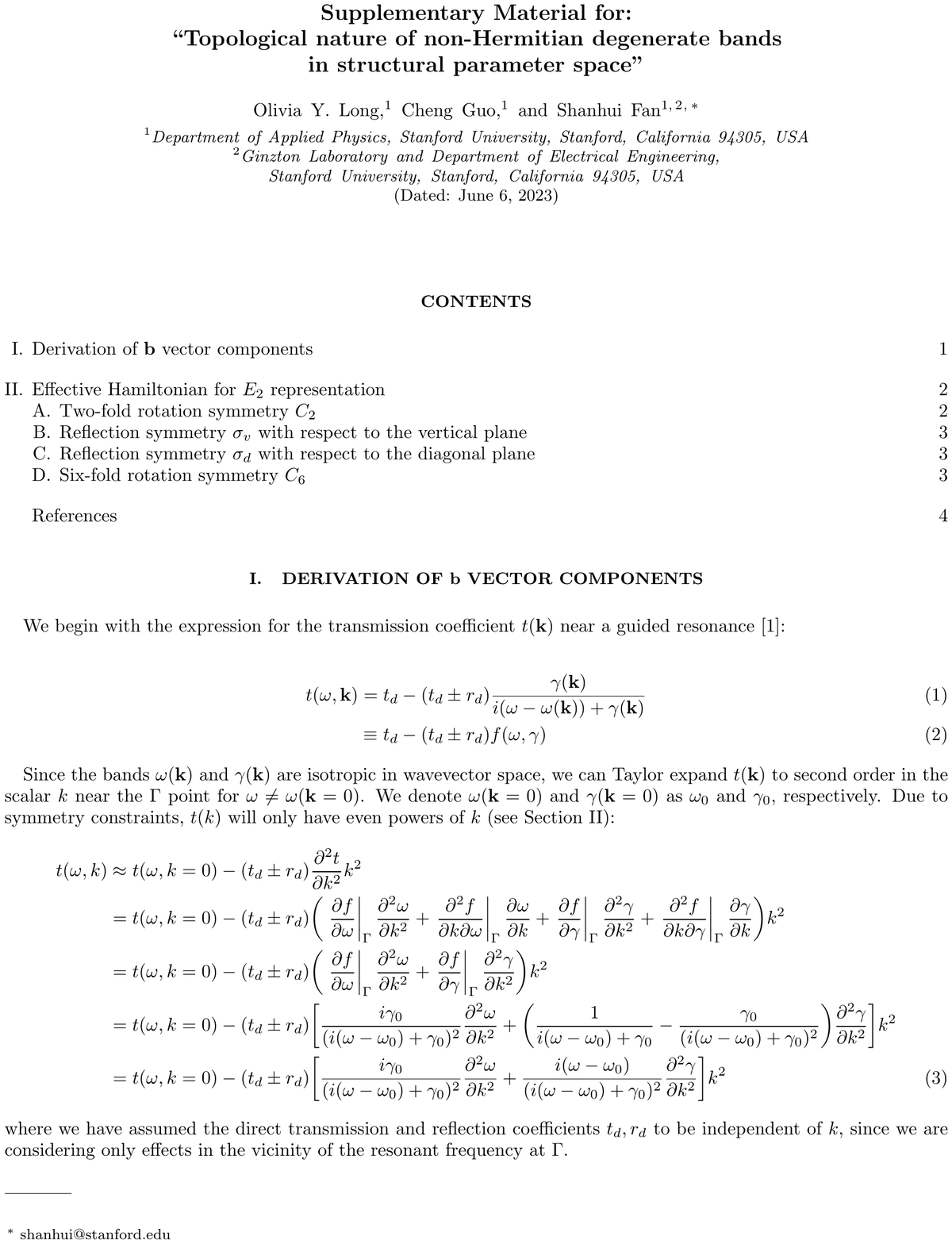}

\pdfximage{\supplementfilename}
\def\numbersupplementpages{\the\pdflastximagepages}

\newif\ifarXiv
\arXivtrue 

\begin{document}

\title{Topological nature of non-Hermitian degenerate bands \\
in structural parameter space 
}

\author{Olivia Y. Long}
\email{olong@stanford.edu}
 \affiliation{Department of Applied Physics, Stanford University, Stanford, California 94305, USA}
 
\author{Cheng Guo}%
\affiliation{Department of Applied Physics, Stanford University, Stanford, California 94305, USA}

\author{Shanhui Fan}

\email{shanhui@stanford.edu}
\affiliation{Department of Applied Physics, Stanford University, Stanford, California 94305, USA}
\affiliation{Ginzton Laboratory and Department of Electrical Engineering,
Stanford University, Stanford, California 94305, USA}

\date{\today}

\begin{abstract}
 In photonics, band degeneracies at high-symmetry points in wavevector space have been shown to exhibit rich physical phenomena.
 However, obtaining degenerate bands away from such points is highly nontrivial.
 In this work, we achieve complex band degeneracy in a photonic crystal structure over a region of momentum space. We show that this band degeneracy corresponds to polarization-independent transmission, which can be harnessed for nonlocal metasurface design.
 Moreover, we find that the band degeneracy manifests as a topological singularity in the structural parameter space of the system. 
 Our work highlights
the importance of topological concepts in the design
of polarization-independent photonic structures.
 
 

\end{abstract}

\maketitle



In the developments of energy band theory for Hermitian systems, band degeneracies at isolated points in momentum space have played a significant role. Well-known examples of such band degeneracies include the Dirac \cite{haldane_dirac_points_PRL_2008} and the Weyl points \cite{topolog_photonics_marin_nature_review_2014}. In photonics, such degeneracies have been used to realize rich physical phenomena, including zero-refractive-index metamaterials \cite{momchil_zero-index_BIC_PRL_2018, dirac_cone_accidental_degen_CT_Chan_nature_mat_2011, all-dielectric_zero-index_metamaterial_valentine_2013}, frozen light at degenerate band edges \cite{frozen_light_DBE_PRE_2006, gain_enhancement_phc_at_DBE_PRB_2016}, and topological phase transitions \cite{topolog_photonics_marin_nature_review_2014,accid_degen_topolog_phase_transitions_phc_OE_2016}. Moreover, these degeneracies in Hermitian band theory provide the starting point for exploring notable features such as exceptional rings and contours that are formed when non-Hermitian perturbation is introduced \cite{marin_exceptional_ring_dirac_cones_nature_2015, non-hermitian_perturb_weyl_hamilt_fan_PRB_2018}. 

In contrast to band degeneracies at isolated points, the possibilities of achieving band degeneracy over a \emph{region} of momentum space has received less attention in photonics. Below, for conciseness, we refer to such band degeneracies over a region as \emph{degenerate bands}. In electronic systems, such degenerate bands do occur either for systems without spin-orbit coupling, or for systems with parity--time symmetry due to the Kramers' degeneracy \cite{kramers_degen_theorem_1952}. For photonic systems, generically, spin-orbit coupling is always present and there is no Kramers' degeneracy. Thus, unlike band degeneracies at isolated points, which may be achieved at high-symmetry points with a nontrivial point group, degenerate bands in photonics typically cannot be achieved by symmetry considerations alone, since any region of momentum space typically contains points with only trivial point groups. 

Degenerate bands can be particularly useful in the creation of polarization-independent nonlocal metasurfaces. These metasurfaces aim to control the transmission or reflection of light as a function of in-plane wavevectors and have found applications in optical analog computing, image processing, and augmented reality systems 
\cite{Silva_2014, kwon2018, multifunc_nonlocal_metasurf_PRL_2020, topolog_diff_long_OL_2021, nonlocal_metasurf_eye-tracking_brongersma_nature_2021, high_Q_metasurface_brongersma_nature_2020, capasso_flat_optic_metasurface_review_science_2022, tengfeng_2021, rho_computational_metamaterials_2022}. Typically, these metasurfaces consist of photonic crystal slab structures, and their properties are strongly controlled by the bands of the guided resonances 
\cite{Fan_guided_res_phc_PRB_2002, apps_of_guided_resonances_Fan_APL_2003, review_2d_phc_fano_resonance_fan_2014, cheng_phc_slab_diff_optica_2018, zhou2020flat, wang2020compact, Long_polariz_indep_PRapplied_2022, Bykov_diff_grating_18, Guo_squeeze_optica_2020, kwon2020}. However, 
the properties of these guided resonances are 
typically polarization-dependent, 
which is undesirable in many applications.

Since guided resonances have radiation losses \cite{phc_textbook}, their band structures are complex. Thus, to achieve complete polarization independence in the optical response of nonlocal metasurfaces, it is essential to create degenerate complex bands.
Previous work on polarization-independent metasurfaces have only utilized degeneracy of the real part of the band structure \cite{Long_polariz_indep_PRapplied_2022}. A study incorporating the entire complex band structure is still lacking.




%

In this work, we show that complex degenerate bands can be achieved in nonlocal metasurfaces. Such complex degenerate bands are associated with a topological singularity in the structural parameter space of the system. 
%
We further show that this singularity can be harnessed for polarization-independent applications such as band-reject and high-pass filters, which are important in image processing and optical analog computing.
%
Our work shows the importance of topological concepts in the design of polarization-independent nonlocal metasurfaces. 
%
%
%

\begin{figure}
\includegraphics[width=0.4\textwidth]{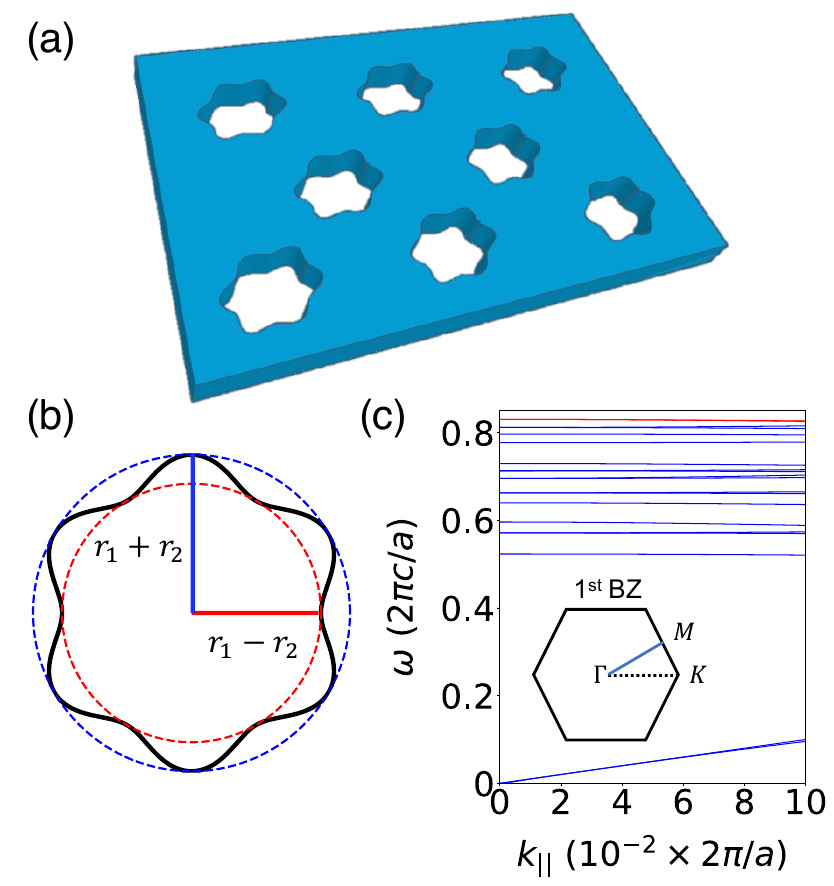}
\caption{\label{structure_fig} (a) Photonic crystal slab with hexagonal array of air holes. The slab has lattice constant $a$ and thickness $d$. The dielectric region of the crystal has relative permittivity $\epsilon$. (b) Shape of air holes described by polar equation: $r(\phi) = r_1 + r_2 \cos{(6\phi)}$. (c) Band structure along $\Gamma K$ direction of a slab with parameters $(r_1, r_2, d, \epsilon) = (0.347 a, -0.033a, 0.2475 a, 13.5)$. Red lines indicate the degenerate bands under study.
}
\end{figure}

To illustrate our theory, we consider a photonic crystal slab possessing $C_{6v}$ symmetry. Our system consists of a slab with daisy-shaped air holes arranged in a hexagonal lattice \cite{momchil_zero-index_BIC_PRL_2018, tang_low-loss_2021}. The slab has lattice constant $a$, thickness $d$, and the dielectric region has a relative permittivity $\epsilon$ (Fig. \ref{structure_fig}a). The air holes are described by the polar equation: $r(\phi) = r_1 + r_2 \cos{(6\phi)}$, as shown in Fig. \ref{structure_fig}b. 

For such a crystal, at normal incidence, light will only couple to a pair of doubly degenerate guided resonance bands at the $\Gamma$ point due to symmetry considerations \cite{Fan_guided_res_phc_PRB_2002, sakoda_hexag_phc_symm_PRB_2001}.
For this crystal with $C_{6v}$ symmetry, the band structure is isotropic in the wavevector space near $\Gamma$, and the $s$ and $p$ polarizations will each couple only to a single band \cite{cheng_phc_slab_diff_optica_2018}. In general, away from the $\Gamma$ point, this pair of bands will split, 
yielding polarization-dependent effects. 


We consider the wavevector-dependent Hamiltonian for our system derived solely from symmetry constraints. 
At the $\Gamma$ point, the system possesses full $C_{6v}$ symmetry. The $C_{6v}$ group supports two different two-dimensional irreducible representations, denoted as $E_1$ and $E_2$ \cite{dresselhaus2007group}, each corresponding to a pair of doubly degenerate modes at $\Gamma$. 
We note that only $E_1$ bands can couple to normally incident light. 


In the vicinity of the $\Gamma$ point, we can derive an effective Hamiltonian using $\mathbf{k} \cdot \mathbf{p}$ theory. 
For cases where the system supports either a pair of $E_1$ or $E_2$ modes at the $\Gamma$ point, the theory yields the following effective Hamiltonian \cite{Guo_squeeze_optica_2020, SM}: 
\begin{align}
    H(\mathbf{k}) &= (\omega_0 - i\gamma_0)\mathbb{I} \nonumber \\
    &+  \begin{pmatrix}
        u|\mathbf{k}|^2 + v(k_x^2 - k_y^2) & 2v k_x k_y \\
        2v k_x k_y & u|\mathbf{k}|^2 - v(k_x^2 - k_y^2)
    \end{pmatrix}
\end{align}
where $u,v \in \mathbb{C}$ and $\omega_0, \gamma_0$ correspond to the resonant frequency and linewidth at the $\Gamma$ point respectively. The parameters $(k_x, k_y)$ are the components of the in-plane wavevector $\mathbf{k}$.
The corresponding dispersion relations are:
\begin{equation}\label{disp_relation}
    \omega_{\pm}(\mathbf{k}) - i\gamma_{\pm}(\mathbf{k}) = \omega_0 -i\gamma_0 + (u \pm v)|\mathbf{k}|^2
\end{equation}
where $+$, $-$ subscripts denote the upper and lower bands, respectively. 

We can label the two bands in Eq. \ref{disp_relation} with the polarization that each band couples to. Let $\omega_{s,p}(\mathbf{k}) = \omega_0 + C_{s,p} \mathbf{k}^2$ and $\gamma_{s,p}(\mathbf{k}) = \gamma_0 + D_{s,p} \mathbf{k}^2$, where $C_{s,p},D_{s,p} \in \mathbb{R}$ are the quadratic coefficients of the real and imaginary band dispersions. Generally, $C_s \neq C_p$ and $D_s \neq D_p$. 
However, when $C_s = C_p$ and $D_s = D_p$, we can create complex degenerate bands, which provide a polarization-independent resonant response over a range of wavevectors. 

We undertake a numerical optimization to achieve such complex degenerate bands. 
For this purpose, it is useful to provide a quantitative measure of how a given band structure deviates from being completely degenerate. Since some of the applications of these structures concern the control of the transmission of light, we relate the differences in the complex band structure to differences in the transmission coefficients of each polarization. 
We consider the transmission coefficients $t_{\sigma \mu}$, where $\mu$ and $\sigma$ denote the polarizations of the incident and transmitted light, respectively. 
Only $t_{ss}(\mathbf{k})$ and $t_{pp}(\mathbf{k})$ are non-zero, since there is no polarization conversion between the $s$ and $p$ polarizations \cite{cheng_phc_slab_diff_optica_2018}.

At an operating frequency $\omega$, the transmission coefficient for polarization $\mu\in \{s,p\}$ as a function of $\mathbf{k}$ is \cite{Fan_guided_res_phc_PRB_2002}:
\begin{equation} \label{transmission_eqn}
    t_{\mu \mu}(\omega, \mathbf{k}) = t_d - (t_d \pm r_d)\frac{\gamma_\mu(\mathbf{k})}{i(\omega - \omega_\mu(\mathbf{k})) + \gamma_\mu(\mathbf{k})}
\end{equation}
where $t_d, r_d$ are the direct transmission and reflection coefficients, which are assumed to be independent of frequency and wavevector, since we are considering only effects in the vicinity of the resonant frequency at $\Gamma$. 
The plus or minus sign in Eq. \ref{transmission_eqn} corresponds to even or odd modes with respect to the plane of the slab.
Near the $\Gamma$ point, we expand $t_{\mu \mu}(\omega, \mathbf{k})$ to second order with respect to $k$ for a general $\omega$:
\begin{align}\label{taylor_expansion_main}
    t_{\mu \mu}(\omega, \mathbf{k}) &\approx t(\omega, \mathbf{k}=0) \nonumber \\
    &-  \frac{i \left.(t_d \pm r_d) \right|_{\Gamma}}{(i(\omega-\omega_0) +\gamma_0)^2} \bigg[ \gamma_0C_{\mu} + (\omega-\omega_0) D_{\mu} \bigg] k^2
\end{align}
%
From Eq. \ref{taylor_expansion_main}, 
%
we define a vector $\mathbf{b}$ as a measure of how differences in the real and imaginary parts of the band curvatures contribute to the polarization dependency $t_{ss}(\mathbf{k}) - t_{pp}(\mathbf{k})$ of the system \cite{SM}:
%
\begin{align}\label{b_vector_def}
\mathbf{b} = \begin{pmatrix}  \gamma_0(C_s - C_p) \\[1em]
(\omega-\omega_0) (D_s-D_p) 
\end{pmatrix}
\end{align}
%
For any frequency $\omega \neq \omega_0$, 
$\mathbf{b}=0$ results in $t_{ss}(\mathbf{k}) = t_{pp}(\mathbf{k})$ to the order of $k^2$. When $\omega = \omega_0$, the effects of $D_\mu$ enter as higher-order $k$ terms. 


\begin{figure}[t]
\includegraphics[width=0.45\textwidth]{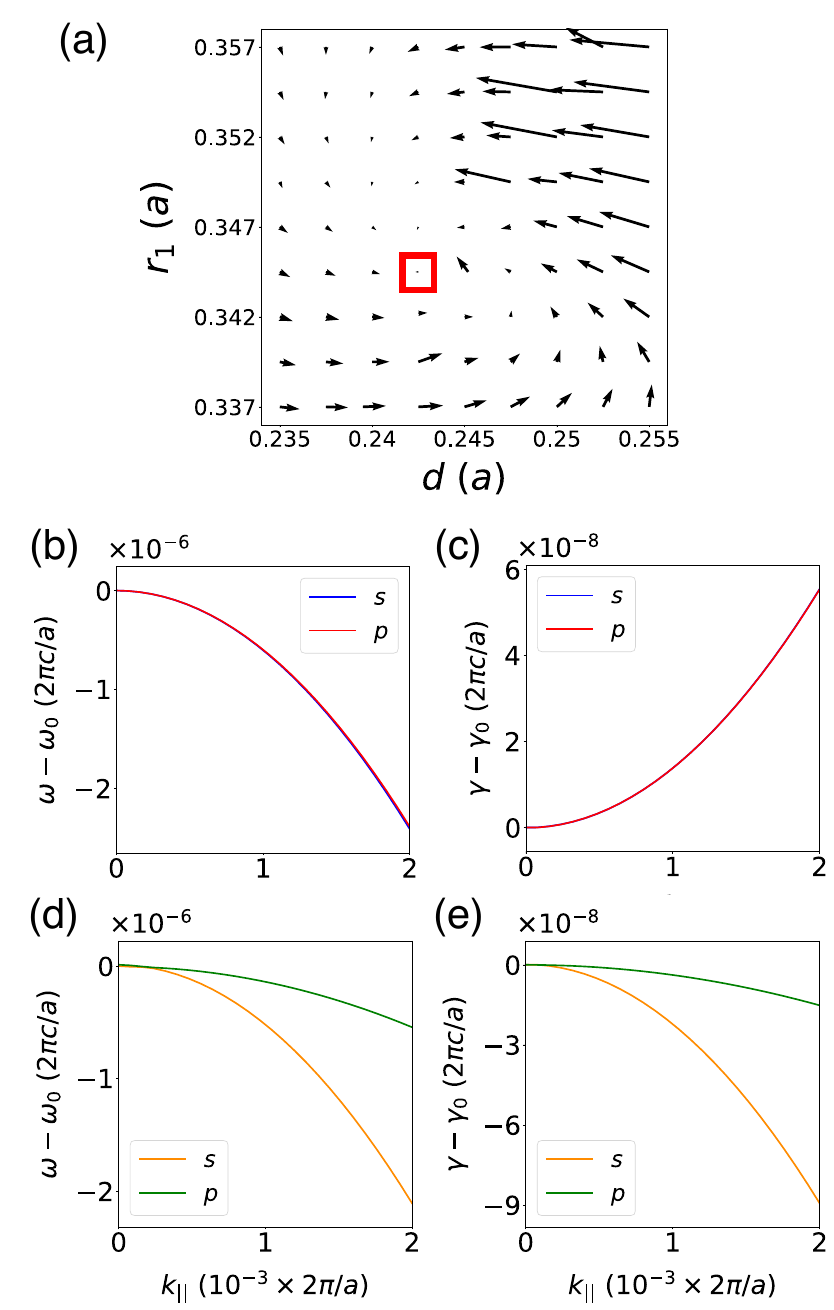}
\caption{\label{vector_plot_fig} (a) Vector field plot of complex $\mathbf{b}$ vector defined in Eq. \ref{b_vector_def} showing the overall $-1$ topological charge around the singularity (red box). The parameters $r_2 = -0.033 a$ and $\epsilon = 13.5$ are held constant, and $\omega-\omega_0 = 0.00106 \times 2\pi c/a$. (b) Band diagram of structure at topological singularity $(r_1, d) = (0.3445 a, 0.2425 a)$ for the real part $\omega(\mathbf{k})$ and (c) imaginary part $\gamma(\mathbf{k})$. (d) Band diagram of structure at $(r_1, r_2, d, \epsilon) = (0.33a, 0.15a, 0.34a, 13.5)$, away from the topological singularity, for the real part and (e) imaginary part. 
}
\end{figure}

\begin{figure}
\includegraphics[width=0.4\textwidth] {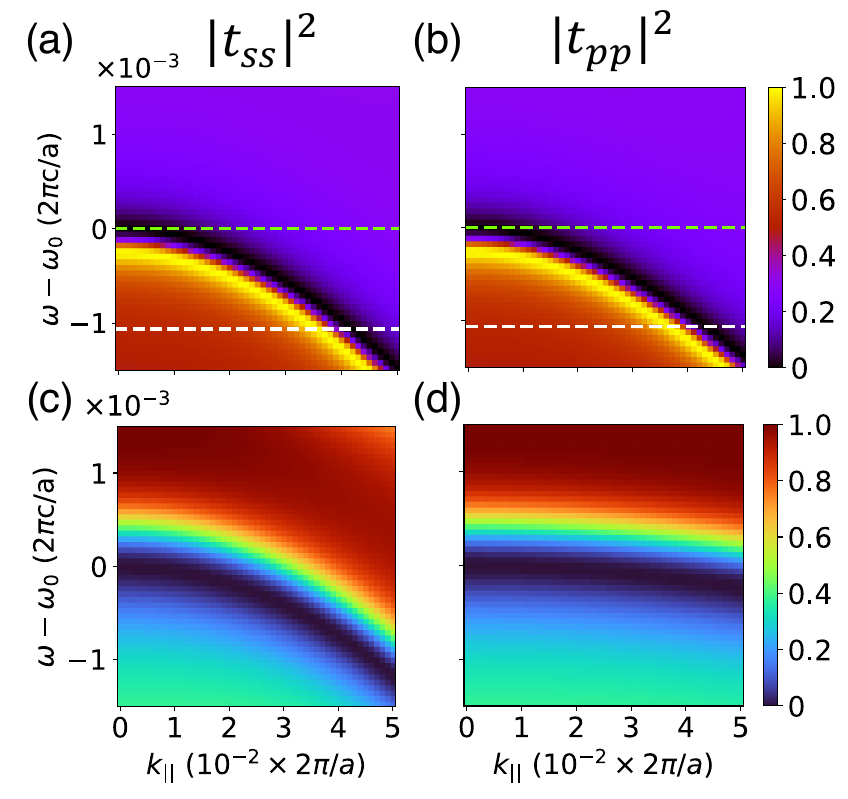}
\caption{\label{RCWA_plots_fig} (a)--(b) Transmission spectra as a function of $k_{||}$ and $\omega-\omega_0$ for $s$ and $p$ polarizations at the topological singularity shown in Fig. \ref{vector_plot_fig}. The dashed green and white lines indicate the operating frequencies of the high-pass and band-reject filters, respectively. (c)--(d) Transmission spectra as a function of $k_{||}$ and $\omega-\omega_0$ for $s$ and $p$ polarizations for a structure away from the topological singularity with parameters $(r_1, r_2, d, \epsilon) = (0.33a, 0.15a, 0.34a, 13.5)$. Plots were computed using the rigorous coupled-wave analysis (RCWA) method \cite{Jin_RCWA_2020}.
}
\end{figure}

\begin{figure*}
\includegraphics[width=0.9\textwidth]{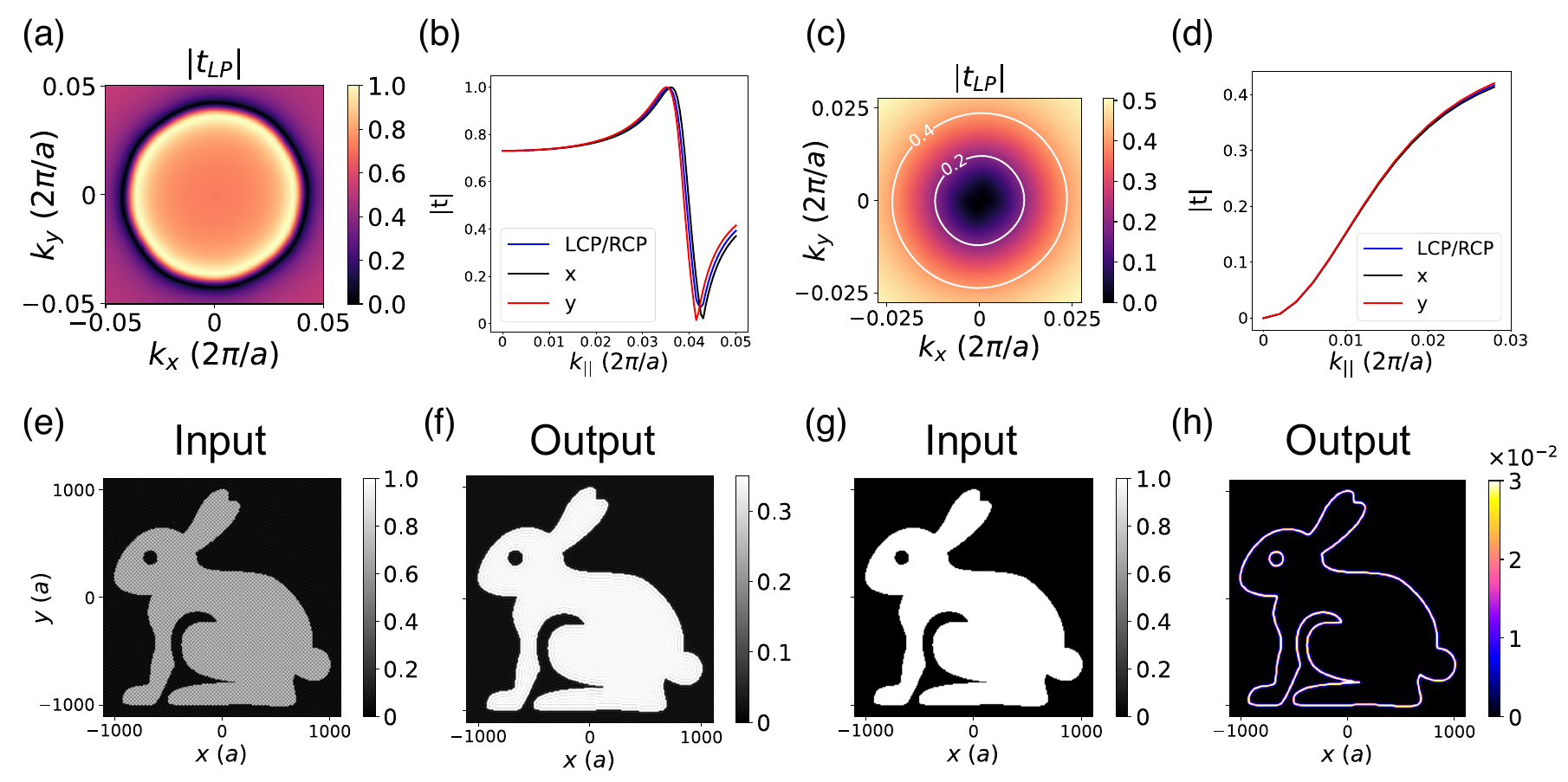}
\caption{\label{num_demo_fig} Numerical demonstration of isotropic spatial frequency filters using complex band degeneracy. (a) Transfer function of band-reject filter at operating frequency $\omega = 0.8305 \times 2\pi c/a$ for $45^{\circ}$ linearly polarized light. (b) Magnitude of band-reject filter transfer function as function of $k_{||}$ for different polarizations. (c) Transfer function at operating frequency $\omega_0 = 0.83167 \times 2\pi c/a$ for $45^{\circ}$ linearly polarized light. (d) Magnitude of high-pass filter transfer function as function of $k_{||}$ for different polarizations. (e)--(f) Intensity profile of linearly polarized input image and transmitted output image for band-reject filter. (g)--(h) Intensity profile of linearly polarized input image and  transmitted output image for high-pass filter.
}
\end{figure*}


For a two-dimensional vector field in a two-dimensional parameter space, any generic isolated singularity will have an associated winding number \cite{roe_winding_around_2015}. Thus, 
we study the topological nature of a singularity in the $\mathbf{b}$ field, which corresponds to the point of complex band degeneracy, in the $(r_1, d)$ structural parameter space. The remaining parameters are fixed: $r_2 = -0.033a$, and $\epsilon = 13.5$, which approximates Si at optical frequencies. 
Each point in the parameter space corresponds to a different photonic crystal slab structure. We choose to study an $E_1$ band near the $\Gamma$ point, delineated by the red bands in Fig. \ref{structure_fig}c, which shows the band structure along the $\Gamma K$ direction for the photonic crystal slab with parameters $(r_1, d) = (0.347a, 0.2475a)$. Here, the resonant frequency at the $\Gamma$ point is $\omega_0 = 0.82606 \times 2\pi c/a$. 
Note that the mode under study only exhibits zeroth-order diffraction. 
%

As we tune the $(r_1, d)$ parameters of our structure, we track the same $E_1$ mode and fit the band structures to a quadratic function to obtain the band curvatures $C_s, C_p, D_s,$ and $D_p$. The band structures were calculated using the guided-mode expansion method (GME) \cite{Minkov_legume_2020, andreani_GME_PRB_2006}. We set $\omega - \omega_0 = 0.00106 \times 2\pi c/a$, which corresponds to the operating frequency of a filter in the wavevector domain, as discussed below. Using this value of $\omega - \omega_0$ and the linewidths $\gamma_0$ at each point in the parameter space, we employ Eq. \ref{b_vector_def} to generate the $\mathbf{b}$ vector field shown in Fig. \ref{vector_plot_fig}a. 
The computed topological singularity is located at $(r_1, d) = (0.3445 a, 0.2425 a)$, where the $\mathbf{b}$ vector magnitude is minimized (red box in Fig. \ref{vector_plot_fig}a).
From the vector plot, we see that the singularity exhibits a topological charge of $-1$. 












In Fig. \ref{vector_plot_fig}b--c, we plot the real and imaginary parts of the complex band structure at the topological singularity along the $\Gamma K$ direction. Note that away from the $\Gamma$ point, both real and imaginary parts of the frequency are indeed degenerate. 
We contrast the band structure at the topological singularity with that of another 
%
photonic crystal slab structure with parameters $(r_1, r_2, d, \epsilon) = (0.33a, 0.15a, 0.34a, 13.5)$.
%
Note that the dispersion relations $\omega(\mathbf{k})$ and $\gamma(\mathbf{k})$ for the real and imaginary parts of the frequency are not degenerate away from $\mathbf{k} = 0$. In this case, $C_s < C_p$ and $D_s < D_p$. 

The degeneracy of the complex band structure away from the $\Gamma$ point should translate to the degeneracy of the transmission coefficients in a range of wavevectors $\mathbf{k}$, i.e. $t_{ss}(\mathbf{k}) = t_{pp}(\mathbf{k})$ for a range of $\mathbf{k}$. To illustrate this, we plot the transmission spectra as a function of the in-plane wavevector $k_{||}$ in Fig. \ref{RCWA_plots_fig}. Panels (a)--(b) show the transmission of the $s$ and $p$ polarizations through the structure corresponding to the topological singularity. At $\mathbf{k}=0$, the transmission spectra exhibits a minimum at the guided resonance frequency $\omega_0 = 0.83167 \times 2\pi c/a$. 
%
%
%
%
%
Away from the $\Gamma$ point, the bands continue to remain degenerate, in agreement with the band structure shown in Fig. \ref{vector_plot_fig}b--c. Note that the transmission spectra also exhibit similar linewidths away from the $\Gamma$ point, in accordance with the degenerate imaginary band frequencies. Thus, the transmittance of this structure 
is polarization-independent. 

In direct contrast, 
in Fig. \ref{RCWA_plots_fig}c--d, we plot the transmission spectra as a function of $k_{||}$ for the same structure as in Fig. \ref{vector_plot_fig}d--e, which is away from the topological singularity. We observe that $C_s < C_p$ and 
$D_s < D_p$. Therefore, the band excitations for this mode are polarization-dependent. The results illustrate the connection between the topological singularity and the polarization independence of the photonic slab structure over a range of frequencies and wavevectors. 



We now apply the complex band degeneracy to wavevector domain filtering, which is useful for image processing \cite{gonzalez_digital_image_book}. Specifically, we demonstrate isotropic high-pass and band-reject filters, which attenuate low frequencies and frequency bands, respectively. Such filters find applications in periodic noise suppression and edge detection \cite{gonzalez_digital_image_book, marr1980theory}.
Previous photonic crystal designs for this functionality yield strong polarization dependence, which can be undesirable in practical applications \cite{isotropic_wavevector_image_filters_cheng_JOSA_2018}. 

To realize the band-reject filter,
we choose the operating frequency to be at the frequency $\omega = 0.8305 \times 2\pi c/a$
as depicted by the dashed white line in Fig. \ref{RCWA_plots_fig}a--b. This corresponds to $\omega-\omega_0 = -0.00106 \times 2\pi c/a$, which is equal in magnitude to the value used in Fig. \ref{vector_plot_fig}.
In Fig. \ref{num_demo_fig}a, we plot the transfer function of the band-reject filter for $45^{\circ}$ linearly polarized light, 
which has an $s$ and $p$ polarization decomposition that depends on the in-plane wavevector $\mathbf{k}$. 
Despite this $\mathbf{k}$-dependent mixture of polarizations, the transfer function for linearly polarized light remains isotropic and thus, polarization-independent. We compare the magnitude of the transmission coefficient as a function of $k_{||}$ for different polarizations in Fig. \ref{num_demo_fig}b, further confirming the polarization independence.

Although the $\mathbf{b}$ vector is defined by a specific $\omega$ value (Eq. \ref{b_vector_def}), when $\mathbf{b} = 0$, the bands are degenerate and can thus be used to realize a high-pass filter at a different operating frequency. Specifically, we operate at the resonant frequency $\omega_0 = 0.83167 \times 2\pi c/a$, depicted by the dashed green line in Fig. \ref{RCWA_plots_fig}a--b. As in the case of the band-reject filter, the transfer function of the high-pass filter for $45^{\circ}$ linearly polarized light is isotropic and polarization-independent (Fig. \ref{num_demo_fig}c). This is confirmed in the plot of the transmission coefficient magnitude as a function of $k_{||}$ for different polarizations (Fig. \ref{num_demo_fig}d).
Using the transfer functions for $45^{\circ}$ linearly polarized light, 
we demonstrate the filter functionalities on an input image of size $2000a \times 2000a$. To generate the transmitted output images, we perform a spatial Fourier transform on the input image, multiply the Fourier transform by the transfer function, then perform the inverse Fourier transform. We utilize the wavevector range $|\mathbf{k}| = [0, 0.05] \times 2\pi /a$ to demonstrate both functionalities.
For the band-reject filter, the input image is corrupted by 2D periodic noise 
(Fig. \ref{num_demo_fig}e). Upon transmission through the structure, the noise is eliminated in the output image, demonstrating isotropic and polarization-independent operation (Fig. \ref{num_demo_fig}f). By changing the operating frequency, the filter can be tuned to reject any desired frequency band. 
To demonstrate the high-pass filter, we utilize the input image shown in Fig. \ref{num_demo_fig}g.
The transmitted output image shows that edges along all directions are detected, confirming the polarization independence of the device (Fig. \ref{num_demo_fig}h).

In summary, we have shown that complex band degeneracy can be achieved over a region of wavevector space in a photonic crystal system. This degeneracy manifests as a topological singularity in the structural parameter space and can be harnessed for polarization-independent applications such as spatial frequency filters for image processing and optical analog computing.
Our work highlights the importance of topological concepts in the design of polarization-independent structures.

\begin{acknowledgments}
This work was supported by the Samsung Advanced Institute of Technology (SAIT) of Samsung Electronics, and by a MURI project (Grant No. FA9550-21-1-0312) from the U. S. Air Force Office of Scientific Research.
O.L. acknowledges support from the NSF Graduate Research Fellowship (Grant No. DGE-1656518) and the Stanford Graduate Fellowship. 
\end{acknowledgments}




\bibliography{apssamp}

\ifarXiv
    

\section*{Supplementary material}

# Supplementary Material for: "Topological nature of non-Hermitian degenerate bands in structural parameter space"

Olivia Y. Long,[1] Cheng Guo,[1] and Shanhui Fan[1,2,*]

[1]*Department of Applied Physics, Stanford University, Stanford, California 94305, USA*
[2]*Ginzton Laboratory and Department of Electrical Engineering, Stanford University, Stanford, California 94305, USA*

(Dated: June 6, 2023)

## CONTENTS

## I. DERIVATION OF b VECTOR COMPONENTS

We begin with the expression for the transmission coefficient $t(\mathbf{k})$ near a guided resonance [1]:

$$t(\omega, \mathbf{k}) = t_d - (t_d \pm r_d)\frac{\gamma(\mathbf{k})}{i(\omega - \omega(\mathbf{k})) + \gamma(\mathbf{k})} \quad (1)$$

$$\equiv t_d - (t_d \pm r_d) f(\omega, \gamma) \quad (2)$$

Since the bands $\omega(\mathbf{k})$ and $\gamma(\mathbf{k})$ are isotropic in wavevector space, we can Taylor expand $t(\mathbf{k})$ to second order in the scalar $k$ near the $\Gamma$ point for $\omega \neq \omega(\mathbf{k}=0)$. We denote $\omega(\mathbf{k}=0)$ and $\gamma(\mathbf{k}=0)$ as $\omega_0$ and $\gamma_0$, respectively. Due to symmetry constraints, $t(k)$ will only have even powers of $k$ (see Section II):

$$\begin{aligned}
t(\omega, k) &\approx t(\omega, k=0) - (t_d \pm r_d)\frac{\partial^2 t}{\partial k^2} k^2 \\
&= t(\omega, k=0) - (t_d \pm r_d)\left(\left.\frac{\partial f}{\partial \omega}\right|_\Gamma \frac{\partial^2 \omega}{\partial k^2} + \left.\frac{\partial^2 f}{\partial k \partial \omega}\right|_\Gamma \frac{\partial \omega}{\partial k} + \left.\frac{\partial f}{\partial \gamma}\right|_\Gamma \frac{\partial^2 \gamma}{\partial k^2} + \left.\frac{\partial^2 f}{\partial k \partial \gamma}\right|_\Gamma \frac{\partial \gamma}{\partial k}\right) k^2 \\
&= t(\omega, k=0) - (t_d \pm r_d)\left(\left.\frac{\partial f}{\partial \omega}\right|_\Gamma \frac{\partial^2 \omega}{\partial k^2} + \left.\frac{\partial f}{\partial \gamma}\right|_\Gamma \frac{\partial^2 \gamma}{\partial k^2}\right) k^2 \\
&= t(\omega, k=0) - (t_d \pm r_d)\left[\frac{i\gamma_0}{(i(\omega - \omega_0) + \gamma_0)^2}\frac{\partial^2 \omega}{\partial k^2} + \left(\frac{1}{i(\omega - \omega_0) + \gamma_0} - \frac{\gamma_0}{(i(\omega - \omega_0) + \gamma_0)^2}\right)\frac{\partial^2 \gamma}{\partial k^2}\right] k^2 \\
&= t(\omega, k=0) - (t_d \pm r_d)\left[\frac{i\gamma_0}{(i(\omega - \omega_0) + \gamma_0)^2}\frac{\partial^2 \omega}{\partial k^2} + \frac{i(\omega - \omega_0)}{(i(\omega - \omega_0) + \gamma_0)^2}\frac{\partial^2 \gamma}{\partial k^2}\right] k^2 \quad (3)
\end{aligned}$$

where we have assumed the direct transmission and reflection coefficients $t_d, r_d$ to be independent of $k$, since we are considering only effects in the vicinity of the resonant frequency at $\Gamma$.

---

[*] shanhui@stanford.edu

For a given polarization $\mu \in \{s, p\}$, Eq. 3 can be expressed as:

$$t_{\mu\mu}(\omega, k) \approx t(\omega, k=0) - (t_d \pm r_d)\left[\frac{i\gamma_0}{(i(\omega-\omega_0)+\gamma_0)^2}C_\mu + \frac{i(\omega-\omega_0)}{(i(\omega-\omega_0)+\gamma_0)^2}D_\mu\right]k^2 \quad (4)$$

where we have used the variables $C_\mu, D_\mu$ to denote the curvatures of the real and imaginary bands respectively, as in the main text.

To achieve complete polarization independence in transmission, we would like to minimize $t_{ss}(k) - t_{pp}(k)$. Note that the polarization conversion terms $t_{sp}(k) = 0$ and $t_{ps}(k) = 0$ due to the single-band excitation phenomenon resulting from an isotropic band structure [2]. Thus, the figure of merit we wish to minimize is:

$$t_{ss}(k) - t_{pp}(k) \approx -(t_d \pm r_d)\left[\frac{i\gamma_0}{(i(\omega-\omega_0)+\gamma_0)^2}(C_s - C_p) + \frac{i(\omega-\omega_0)}{(i(\omega-\omega_0)+\gamma_0)^2}(D_s - D_p)\right]k^2$$

$$= -(t_d \pm r_d)\frac{i}{(i(\omega-\omega_0)+\gamma_0)^2}\left[\gamma_0(C_s - C_p) + (\omega-\omega_0)(D_s - D_p)\right]k^2 \quad (5)$$

Using the two terms above, we define the **b** vector quantity as a measure of the polarization dependence of the system:

$$\mathbf{b} = \begin{pmatrix} b_1 \\ b_2 \end{pmatrix} = \begin{pmatrix} \gamma_0(C_s - C_p) \\ (\omega-\omega_0)(D_s - D_p) \end{pmatrix} \quad (6)$$

From Eq. 6, we see that $\mathbf{b} = 0$ corresponds to when $t_{ss}(k) = t_{pp}(k)$.

## II. EFFECTIVE HAMILTONIAN FOR $E_2$ REPRESENTATION

In this section, we consider a photonic crystal slab with $C_{6v}$ symmetry as shown in Fig. 1 of the main text, and derive the functional form of the dispersion relation near the $\Gamma$ point for the $E_2$ mode. The derivation for the $E_1$ mode can be found in the Supplementary of Ref. [3].

The general form of the Hamiltonian is:

$$H(\mathbf{k}) = A(\mathbf{k}) - iB(\mathbf{k}) \quad (7)$$

where $A(\mathbf{k})$ and $B(\mathbf{k})$ are $2 \times 2$ Hermitian matrices, and $\mathbf{k} = (k_x, k_y)$ is the in-plane wavevector. We can write the matrices in the general form:

$$A(\mathbf{k}) = \begin{pmatrix} \omega_0 + h(\mathbf{k}) + g(\mathbf{k}) & f(\mathbf{k}) \\ f^*(\mathbf{k}) & \omega_0 + h(\mathbf{k}) - g(\mathbf{k}) \end{pmatrix} \quad (8)$$

$$B(\mathbf{k}) = \begin{pmatrix} \gamma_0 + t(\mathbf{k}) + s(\mathbf{k}) & r(\mathbf{k}) \\ r^*(\mathbf{k}) & \gamma_0 + t(\mathbf{k}) - s(\mathbf{k}) \end{pmatrix} \quad (9)$$

where $h, g, t, s \in \mathbb{R}$ and $f, r \in \mathbb{C}$ since $A, B$ are Hermitian matrices.

The symmetry transformations of the Hamiltonian obey the following relation:

$$D(\mathbf{k})H(\mathbf{k})D(\mathbf{k})^\dagger = H(\mathcal{D}(\mathbf{k})) \quad (10)$$

where $D, \mathcal{D}$ are the representations of the symmetry transformations in the Hilbert space of the modes and the wavevector space, respectively. We use the basis functions $|x\rangle$ and $|y\rangle$ for our matrix representations.

### A. Two-fold rotation symmetry $C_2$

The $E_2$ representation of the two-fold rotation operation $C_2$ is:

$$D(C_2) = \begin{pmatrix} 1 & 0 \\ 0 & 1 \end{pmatrix} \quad (11)$$

Since this is just the identity matrix, we do not obtain any constraints on $H$ using this symmetry.

## B. Reflection symmetry $\sigma_v$ with respect to the vertical plane

The $E_2$ representation for the reflection operation $\sigma_v$ with respect to the vertical plane is:

$$D(\sigma_v) = \begin{pmatrix} 1 & 0 \\ 0 & -1 \end{pmatrix} \tag{12}$$

Applying this symmetry transformation to Eqn. 10:

$$\begin{pmatrix} h(k_x, k_y) + g(k_x, k_y) & -f(k_x, k_y) \\ -f^*(k_x, k_y) & h(k_x, k_y) - g(k_x, k_y) \end{pmatrix} = \begin{pmatrix} h(k_x, -k_y) + g(k_x, -k_y) & f(k_x, -k_y) \\ f^*(k_x, -k_y) & h(k_x, -k_y) - g(k_x, -k_y) \end{pmatrix} \tag{13}$$

The general form of $f(\mathbf{k})$ is: $f(k_x, k_y) = f_x k_x + f_y k_y + f_{xx} k_x^2 + f_{yy} k_y^2 + f_{xy} k_x k_y$. To satisfy the above, we get: $f_x = 0, f_{xx} = f_{yy} = 0$. Similarly, the constraints on $g(k_x, k_y)$ and $h(k_x, k_y)$ are: $g_y = 0, g_{xy} = 0$ and $h_y = 0, h_{xy} = 0$, respectively.

With the constraints from $\sigma_v$, the functions $f$, $g$ and $h$ take the form:

$$f(k_x, k_y) = f_y k_y + f_{xy} k_x k_y \tag{14}$$
$$g(k_x, k_y) = g_x k_x + g_{xx} k_x^2 + g_{yy} k_y^2 \tag{15}$$
$$h(k_x, k_y) = h_x k_x + h_{xx} k_x^2 + h_{yy} k_y^2 \tag{16}$$

## C. Reflection symmetry $\sigma_d$ with respect to the diagonal plane

The $E_2$ representation for the reflection operation $\sigma_d$ with respect to the diagonal plane is:

$$D(\sigma_d) = \begin{pmatrix} -1 & 0 \\ 0 & 1 \end{pmatrix} \tag{17}$$

Applying this symmetry transformation to Eqn. 10:

$$\begin{pmatrix} h(k_x, k_y) + g(k_x, k_y) & -f(k_x, k_y) \\ -f^*(k_x, k_y) & h(k_x, k_y) - g(k_x, k_y) \end{pmatrix} = \begin{pmatrix} h(-k_x, k_y) + g(-k_x, k_y) & f(-k_x, k_y) \\ f^*(-k_x, k_y) & h(-k_x, k_y) - g(-k_x, k_y) \end{pmatrix} \tag{18}$$

Using the current forms of $f, g, h$ (Eq. 14–16), we find:

$$h(k_x, k_y) = h(-k_x, k_y) \to h_x = 0 \tag{19}$$
$$g(k_x, k_y) = g(-k_x, k_y) \to g_x = 0 \tag{20}$$
$$-f(k_x, k_y) = f(-k_x, k_y) \to f_y = 0 \tag{21}$$

Now, with the constraints from $\sigma_d$, the functions $f$, $g$ and $h$ take the form:

$$f(k_x, k_y) = f_{xy} k_x k_y \tag{22}$$
$$g(k_x, k_y) = g_{xx} k_x^2 + g_{yy} k_y^2 \tag{23}$$
$$h(k_x, k_y) = h_{xx} k_x^2 + h_{yy} k_y^2 \tag{24}$$

## D. Six-fold rotation symmetry $C_6$

The $E_2$ representation for the six-fold rotation operation $C_6$ is:

$$D(C_6) = \begin{pmatrix} -\frac{1}{2} & -\frac{\sqrt{3}}{2} \\ \frac{\sqrt{3}}{2} & -\frac{1}{2} \end{pmatrix} \tag{25}$$

Applying this symmetry transformation to Eqn. 10, we get the equality:

$$\begin{pmatrix} h(k_x, k_y) - \frac{1}{2} g(k_x, k_y) + \frac{\sqrt{3}}{2} \operatorname{Re}\{f\} & -\frac{\sqrt{3}}{2} g + \frac{i}{2} \operatorname{Im}\{f\} - \frac{1}{2} f^* \\ -\frac{\sqrt{3}}{2} g - \frac{i}{2} \operatorname{Im}\{f\} - \frac{1}{2} f & h(k_x, k_y) + \frac{1}{2} g(k_x, k_y) - \frac{\sqrt{3}}{2} \operatorname{Re}\{f\} \end{pmatrix} = \begin{pmatrix} h(D(k)) + g(D(k)) & f(D(k)) \\ f^*(D(k)) & h(D(k)) - g(D(k)) \end{pmatrix} \tag{26}$$

For the $h(k)$ terms, we get:

$$h_{xx}k_x^2 + h_{yy}k_y^2 = h_{xx}\left(-\frac{1}{2}k_x - \frac{\sqrt{3}}{2}k_y\right)^2 + h_{yy}\left(\frac{\sqrt{3}}{2}k_x - \frac{1}{2}k_y\right)^2 \tag{27}$$

$$\to h_{xx}k_x^2 = \left(\frac{1}{4}h_{xx} + \frac{3}{4}h_{yy}\right)k_x^2 \tag{28}$$

$$\to h_{xx} = h_{yy} \tag{29}$$

For the $g(k)$ terms in the diagonal of the matrix, we get:

$$-\frac{1}{2}(g_{xx}k_x^2 + g_{yy}k_y^2) = g_{xx}\left(-\frac{1}{2}k_x - \frac{\sqrt{3}}{2}k_y\right)^2 + g_{yy}\left(\frac{\sqrt{3}}{2}k_x - \frac{1}{2}k_y\right)^2 \tag{30}$$

$$\to -\frac{1}{2}g_{xx}k_x^2 = \frac{1}{4}g_{xx}k_x^2 + \frac{3}{4}g_{yy}k_x^2 \tag{31}$$

$$\to g_{yy} = -g_{xx} \tag{32}$$

Finally, in the off-diagonal terms of the matrix, we have:

$$-\frac{\sqrt{3}}{2}g + \frac{i}{2}\text{Im}\{f\} - \frac{1}{2}f^* = f_{xy}\left(-\frac{1}{2}k_x - \frac{\sqrt{3}}{2}k_y\right)\left(\frac{\sqrt{3}}{2}k_x - \frac{1}{2}k_y\right) \tag{33}$$

$$\to -\frac{\sqrt{3}}{2}g_{xx}k_x^2 = f_{xy}\left(-\frac{\sqrt{3}}{4}\right)k_x^2 \tag{34}$$

$$\to g_{xx} = \frac{f_{xy}}{2} \tag{35}$$

Since $g$ is real (because $H$ is Hermitian), $f_{xy}$ is real and thus, $f(k)$ is real and $\text{Im}(f) = 0$.
Thus, we have found that:

$$h_{xx} = h_{yy} \tag{36}$$
$$g_{yy} = -g_{xx} \tag{37}$$
$$g_{xx} = \frac{f_{xy}}{2} \tag{38}$$
$$\text{Im}(f) = 0 \tag{39}$$

The final forms of the functions $f$, $g$ and $h$ are:

$$h(k_x, k_y) = h_{xx}(k_x^2 + k_y^2) \tag{40}$$
$$g(k_x, k_y) = g_{xx}(k_x^2 - k_y^2) \tag{41}$$
$$f(k_x, k_y) = 2g_{xx}k_xk_y \tag{42}$$

The Hermitian matrix $A(\mathbf{k})$ has the form:

$$A(\mathbf{k}) = \begin{pmatrix} (h_{xx} + g_{xx})k_x^2 + (h_{xx} - g_{xx})k_y^2 & 2g_{xx}k_xk_y \\ 2g_{xx}k_xk_y & (h_{xx} - g_{xx})k_x^2 + (h_{xx} + g_{xx})k_y^2 \end{pmatrix} \tag{43}$$

The matrix $B(\mathbf{k})$ has the same form since it obeys the same symmetries.
Since $H(\mathbf{k}) = A(\mathbf{k}) - iB(\mathbf{k})$, the Hamiltonian takes the form:

$$H(\mathbf{k}) = \begin{pmatrix} a|\mathbf{k}|^2 + b(k_x^2 - k_y^2) & 2bk_xk_y \\ 2bk_xk_y & a|\mathbf{k}|^2 - b(k_x^2 - k_y^2) \end{pmatrix} + \begin{pmatrix} \omega_0 - i\gamma_0 & 0 \\ 0 & \omega_0 - i\gamma_0 \end{pmatrix} \tag{44}$$

where $a, b \in \mathbb{C}$.
Diagonalizing the Hamiltonian, we get the following dispersion relation:

$$E(\mathbf{k}) = \omega_0 - i\gamma_0 + (a \pm b)|\mathbf{k}|^2 \tag{45}$$

The final form of this dispersion is isotropic and matches the form resulting from the 2D $E_1$ representation as well.

---


\fi

\end{document}